\documentclass[twocolumn,showpacs,preprintnumbers,amsmath,amssymb]{revtex4}
\usepackage{graphicx}

\newcommand{\ii}{\mathrm{i}}
\newcommand{\dd}{\mathrm{d}}
\newcommand{\ee}{\mathrm{e}}
\newcommand{\bra}[1]{{\langle #1 \vert}}
\newcommand{\ket}[1]{{\vert #1 \rangle}}

\begin{document}

\title{Loschmidt echoes in two-body random matrix ensembles}
\author{Iztok Pi{\v z}orn}
\affiliation{Physics Department, Faculty of mathematics and physics, University of Ljubljana,
Jadranska 19, SI-1000 Ljubljana, Slovenia}

\author{Toma{\v z} Prosen}
\affiliation{Physics Department, Faculty of mathematics and physics, University of Ljubljana,
Jadranska 19, SI-1000 Ljubljana, Slovenia}

\author{Thomas H. Seligman}
\affiliation{Centro Internacional de Ciencias, Apartado postal 6-101, C.P.62132 Cuernavaca, Morelos, Mexico}
\affiliation{Instituto de Ciencias F\'\i sicas, Universidad Nacional Aut\'onoma de M\'exico, C.P.62132 Cuernavaca, Morelos, Mexico}

\begin{abstract}
Fidelity decay is studied for quantum many-body systems with a dominant
independent particle Hamiltonian resulting e.g. from a mean field theory
with a weak two-body interaction.  The diagonal terms of the interaction
are included in the  unperturbed Hamiltonian, while the off-diagonal
terms constitute the perturbation that distorts the echo. We give the
linear response solution for this problem in a random matrix framework.
While the ensemble average shows no surprising behavior, we find that the
typical ensemble member as represented by the median displays 
a very slow fidelity decay known as ``freeze''. Numerical
calculations confirm this result and show that the ground state even on average
displays the freeze. This may contribute to explanation of the ``unreasonable'' 
success of mean field theories.
\end{abstract}


\maketitle

\section{Introduction}
Wigner \cite{wigner1} proposed the use of random matrix theory, in particular of
the classical ensembles \cite{cartan, brody} to explain statistical properties of
nuclear spectra. Balian \cite{balian} showed that this amounts to
using a minimum information approach, where symmetry is the only characteristics
taken into account. An essential element for the widespread success \cite{guhr}
resides in the ergodicity of the results obtained for classical ensembles.
Yet it was soon noticed \cite{frenchPL33B,bohigasPL34B}
that the essential two-body character of the underlying interaction requires embedded
ensembles, in particular the two-body random matrix ensemble (TBRE).
While the original work carried the full weight of a three dimensional
few nucleon system, soon abstract models with structureless fermions
were introduced \cite{french2}
to understand more easily embedded
ensembles in general and the TBRE  in particular. Two problems beset
these studies: one is the apparent non-ergodicity of these ensembles,
the other is the difficulty to obtain any analytical results
(see \cite{benet} for a review). After individual
unfolding of each spectrum \cite{brody,bohigasPL34B} spectral statistics
of the classical ensembles are recovered and this marginated the interest
in embedded ensembles for some time. Interest was rekindled
recently mainly by studies of interesting properties at the edge
of the spectra concerning ground states \cite{bertsch1, papenbrockweiden,frank}.
The question whether the ensemble is actually ergodic or not remains open
\cite{benet}, but French
\cite{french3,benet} has shown that it is somewhat academic, as even the
spectral density has fluctuations of the order $1/\log N$ where $N$ is the
dimension of the Hilbert space. This makes individual unfolding essential
for any statistical analysis of spectra. The reason why individual unfolding
does lead to the right answers is not understood to this day.
The renewed interest in the TBRE has spread to various fields
including applications {\cite{benet, zelevinsky,floresseligman}
and extensions to bosons \cite{kota, benet}.

Fidelity analysis for the stability of quantum systems under perturbation,
first proposed by Peres \cite{peresPRA30}, has become very
fashionable since the advent of
quantum information where it provides a standard criterion \cite{nielsenbook}
of stability;
for a recent review see \cite{fidelityreview}. 
It seems reasonable to check
how a small residual two-body interaction affects 
the stability of the solution of an independent particle Hamiltonian.
The diagonal part of this interaction does not affect the eigenstates
and is usually included in the independent particle term, known as
the mean field approximation.

We focus on the validity of the mean field approximation, given by
the stability of the time-dependent solution of the mean field Hamiltonian 
which is measured by fidelity. This is also the main physical motivation behind
defining the random matrix model in the present paper.

Fidelity has been calculated by the method of stationary phase applied to time-dependent
propagator for semi-classical considerations \cite{jalabert,cucchietti,lewenkopf,vanicek} 
and by super-symmetric techniques \cite{stoeckmannNJP,stoeckmannPRE73} for
a random matrix model presented in \cite{gorinNJP6}. In both contexts most of the 
relevant results can be obtained by perturbative or linear response techniques, which actually do 
not require either model as a basis \cite{P02,PZ02,Progress,gorinNJP6,fidelityreview,G-prl}.
Even in the regime where fidelity is not anymore close to one,
a simple exponentiation of the leading term yields excellent results far beyond this regime 
\cite{Progress, fidelityreview}.  
Crossovers between semiclassical  and perturbative considerations 
(the latter also known as Fermi Golden rule regime) have been discussed in \cite{jacquodPRE,cerruti}.
A drastic reduction of fidelity decay
termed ``fidelity freeze'' occurs when the perturbation is strictly off diagonal (or more generally,
when its time average is zero) \cite{freezeint,freezech,G-prl}. 
Such a situation can occur when perturbation breaks the antiunitary symmetry (like time-reversal)
of the unperturbed Hamiltonian in an optimal way \cite{G-prl}, which can perhaps be experimentally
realized by some magnetic interactions.

In case of high level fidelity freeze the dynamics is stable for a very long time.
The question is
whether we can see this effect in the situation under consideration. On the one hand
our system looks
promising, due to the inclusion of the diagonal part of the interaction in the
``unperturbed'' independent particle Hamiltonian. On the other hand the occurrence of the freeze
in random matrix models depends heavily on level repulsion implicit in a GOE or a GUE
for the unperturbed system.
Clearly the independent particle model does not have this repulsion and indeed we can expect
results similar to those of a random spectrum where absence of a diagonal perturbation 
to lowest order only suppresses the quadratic decay \cite{fidelityreview}.

Our analysis will show that the ensemble average does not display fidelity freeze.
The typical ensemble member however does show this freeze. This is confirmed
when we consider the median of fidelity rather than its average. This fact is important
by itself as it shows that quantum evolution of the full problem will follow
that of the independent particle problem or mean field approximation 
for a very long time for most systems.
It also is important because it shows that RMT models can describe physical
situations qualitatively, even when the average behavior is completely off.
The last point can also be reformulated as a necessity to look at the right quantities
i.e. quantities whose distributions have no long tails. If we had replaced fidelity
by distortion as introduced in an elastic problem \cite{weaver}, then the average and
median would coincide. This results because distortion is defined as the logarithm of
one minus fidelity \cite{gorin-weaver}.
Indeed in other contexts taking averages of the logarithm is a common
remedy to eliminate exaggerated effects of tails in distributions  \cite{leyvraz-privat}.
Furthermore we will see that for the ground state and the first excited state of the
independent particle model the freeze will occur even on average.

After introducing basics about fidelity and the Hamiltonian 
we will see that under specific assumptions for the
independent particle spectrum
we can obtain ensemble averaged fidelity decay of a TBRE in
the linear response
regime. The result essentially yields the linear decay. Yet a numerical
inspection of individual Hamiltonians
in the ensemble shows the existence of fidelity freeze.
A more careful numerical analysis shows that the median fidelity
for the ensemble indeed displays the freeze.
Thus we can consider it as typical. This result is quantitatively
reconfirmed by showing that 
the ensemble averaged logarithm of one minus fidelity also displays the freeze.
We relate this behavior directly to the existence of a gap
in  the nearest neighbor spacing distribution of the independent particle spectrum.
Furthermore we find that the fidelity decay of the ground and first excited state
of the independent particle Hamiltonian display the freeze even on average.
Next we numerically check different options for
the single particle spectrum  to obtain a better feeling of the non-ergodicity, which plays
a central role in this context. Finally we give some conclusions.

\section{Fidelity amplitude}

The fidelity amplitude measures the overlap between two quantum states $\psi_0(t')$
and $\psi(t')$ evolving from the same initial state $\psi(0)$
but propagated with slightly different Hamiltonians $H_0$ and $H=H_0+\lambda V$ respectively
\begin{equation}
       f(t') = \langle\psi_0(t')\vert\psi(t')\rangle =%
	   \langle \psi(0)\vert\ee^{\frac{\ii}{\hbar} t' H_0}\ee^{-\frac{\ii}{\hbar} t' H}|\psi(0)\rangle.
\end{equation}

For small perturbation strength $\lambda$ it is convenient to use a linear response
or Born expansion \cite{P02,PZ02, fidelityreview}. We shall follow
the notation established in
\cite{gorinNJP6} as it is well adapted to RMT. In the interaction picture the wave function
reads as
$x(t') = \ee^{\frac{\ii}{\hbar} H_0 t'}\psi(t')$.
Time $t'$ can be replaced by dimensionless time $t$ measured in units of the
Heisenberg time $t_H = 2\pi\hbar/d$ where $d$ denotes the average level spacing
in the spectra of $H_0$ which can and will be set to one.
Therefore, the time will always be given in units of $t_H$.
Time evolution of a state $x(t)$ up to the second order $x^{(2)}(t)=X(t)x(0)$
in perturbation strength $\lambda$ can be expressed as
\begin{eqnarray}
	X(t) &=& 1-2\pi\lambda\ii \int_0^t\dd t_1V_I(t_1)  \nonumber \\
	&-& (2\pi\lambda)^2 \int_0^t\dd t_1 \int_0^{t_1}\dd t_2 V_I(t_1) V_I(t_2)
\label{eq:linresponse}
\end{eqnarray}
where $V_I(t)$ is the abbreviation for the perturbation in the interaction picture
$V_I(t) = \ee^{2\pi\ii H_0} V \ee^{-2\pi\ii H_0}$. 
Please note that the expansion in~(\ref{eq:linresponse}) 
is valid even for times much longer than 
the Heisenberg time ($t\gg 1$), 
if only the perturbation strength is small enough.

The fidelity amplitude for
an initial state $\ket{x(0)}=\sum_j x_\alpha |\alpha\rangle$ written as a superposition
of independent particle states, i.e. state eigenstates of the unperturbed Hamiltonian $\ket{\alpha}$,
then reads
\begin{equation}
       f(t) = \langle x(0) | X(t) | x(0)\rangle =
       \sum_{\alpha,\beta=0}^{\mathcal{N}} X_{\alpha\beta} x_\alpha^* x_\beta.
\end{equation}
Later we consider ensemble averages. Then the linear term
in~(\ref{eq:linresponse}) vanishes and the matrix element $X_{\alpha\beta}$ is
reduced to contributions from the quadratic term only.

\section{The Hamiltonian}

Although a simple RMT of full Gaussian matrices often yields
impressively accurate results, it, strictly speaking, only applies to dynamical systems
where all levels are coupled by the
interaction. More realistic physical systems
such as nuclear or atomic shell models or quantum dots do not possess that property.
The interaction involves two particles only. In the framework of RMT it should be described
by Two-Body Random Ensembles
(TBRE)~\cite{frenchPL33B,bohigasPL34B,brody,zelevinsky,benet}.
We consider a system of $M$ ``spinless''
\footnote{\emph{Spinless} here stands for a description,
where no angular momentum properties are taken into account - different spin states
would correspond to different orbitals, often referred to as spin-orbitals in molecular
physics. This contrasts with the original TBRE \cite{frenchPL33B,bohigasPL34B,brody}
where angular momentum coupling and fractional parentage play an important
role \cite{papenbrockweiden}.}
Fermi particles in $N$ orbitals
in the presence of fermion-fermion interaction. We shall use second quantization notation
with fermion creation and annihilation operators $c_i^\dagger$ and $c_i$ fulfilling the usual
anticommutation relations and creating or annihilating a fermion in the $i$th eigenstate of the single particle
Hamiltonian defined by the mean field. The $M$ particle eigenstate $\ket{\alpha} =
(c_0^\dagger)^{\alpha_0}\cdots (c_{N-1}^\dagger)^{\alpha_{N-1}} \ket{0}$ (where exactly $M$
binary digits $\alpha_i$ of integer $\alpha$ are equal to $1$) of the independent particle
Hamiltonian is then a product of $M$ creation operators with distinct indices applied to a vacuum state.
The Hamiltonian is written in the usual manner as
\begin{equation}
       H = \sum_{i} e_i c_i^\dagger c_i + \lambda \sum_{i<j,k<l} V_{ijkl} c_i^\dagger c_j^\dagger c_l c_k,
       \label{eq:H}
\end{equation}
where all Latin indices are running from $0$ to $N-1$, and
where the $e_i$ are ordered single-particle energies, $V_{ijkl}$ are properly
antisymmetrized two-body matrix elements and $\lambda$ is the perturbation strength.
The unperturbed Hamiltonian $H_0$ will correspond to the Mean Field Approximation and 
will hence be chosen as the one-body terms $\sum_i e_i c_i^\dagger c_i$
plus the diagonal part of the interaction
$\lambda \sum_{i<j} V_{ijij}c_i^\dagger c_j^\dagger c_j c_i$.
A natural question is how the remaining part of $V$, which we shall consider as perturbation, affects the dynamics.

The inclusion of the diagonal part of the interaction in the unperturbed system is crucial to our argument.
On one hand it does not affect the eigenfunctions of the one-particle term and thus it seems of little importance.
Yet in the calculation of fidelity the dephasing it produces, becomes the dominant term. On the other hand, if included
in the unperturbed part it will only enter the result to the fourth order in ${\lambda}$  because the
spectral two-point function will only enter to order $\lambda^2$ and it, 
in turn, will only be affected to order $\lambda^2$ by the diagonal terms.

So far we have not put any constraints on the spectral properties of $H_0$ nor on the
perturbation $V$. The latter consists of independent two-body matrix elements,
where the weight coefficients $V_{ijkl}$ are chosen as
independent random Gaussian complex numbers with vanishing mean and
$\langle V_{ijkl} V_{mnop}\rangle_V = v^{-2} \delta_{ij,op}\delta_{kl,mn}$,
where the variance $v^{-2}$ is set to normalize the perturbation as
$\langle {\rm tr}[V^2] \rangle = \mathcal{N}$ where
$\mathcal{N}={N\choose M}$ is the dimension of full Hilbert space.
The unperturbed dynamics is given by $N$ single-particle levels $e_j$. We shall
often choose them as the eigenvalues of the $N\times N$ Hermitian matrix chosen from the GUE,
but we shall mention other options and their consequences. 
Similar models have recently been widely studied e.g. 
in nuclear physics \cite{flambaum} and for studying chaotic quantum dots \cite{alhassid}.

Two-body operators $c_i^\dagger c_j^\dagger c_l c_k$ can be split into three
groups by inspecting indices $i,j,k,l$: diagonal terms with two pairs of equal indices, 
three-orbital terms with one pair of equal indices, and four-orbital terms without pairs of equal indices.
The non-zero matrix elements
$V_{\alpha\beta}$ in the full Hilbert space therefore couple
many-particle states which differ by at most two single-particle states
and these states are the only intermediate states in a two-fold transition
described by $V_I(t_1) V_I(t_2)$ in~(\ref{eq:linresponse}).
After averaging over an ensemble of two-body interactions
all products of independent $V_{ijkl}$ vanish
and our interest is narrowed to transitions with the same initial and final state
described by averaged matrix elements
$\langle[V(t_1)V(t_2)]_{\alpha\beta}\rangle_V=\delta_{\alpha\beta}C_\alpha(t_1-t_2)$
which only depend on the time difference and are
denoted by the \emph{correlation function} $C_\alpha(\tau)$.
Using $H_0 \equiv {\rm diag}(h_\alpha)$,  which $h_\alpha$ is either
single particle Hamiltonian 
$h_\alpha = \bra{\alpha} \sum_i {\hat n_i} e_i \ket{\alpha}$, ${\hat n_i} = c_i^\dagger c_i$, 
or the mean field Hamiltonian 
$h_\alpha = \bra{\alpha}\sum_i {\hat n_i} e_i + \lambda \sum_{i<j}V_{ijij}{\hat n_i} {\hat n_j} \ket{\alpha}$,
the correlation function can be written as
\begin{eqnarray}
       C_\alpha(\tau) &=& 
       \sum_{\gamma} \langle \ee^{2\pi\ii h_\alpha} V_{\alpha\gamma} \ee^{-2\pi \ii h_\gamma} V_{\gamma\alpha}\rangle_V \nonumber \\
       &=&  \sum_{\gamma} \langle \ee^{-2\pi \ii (h_\gamma-h_\alpha)} \rangle_V \langle V_{\alpha\gamma}  V_{\gamma\alpha}\rangle_V .
\end{eqnarray}
The first average is nontrivial in the mean field case where
the energy differences  $h_\gamma - h_\alpha$ contain also diagonal two-body terms (mean field).
The correction to
$\ee^{-2\pi \ii \sum_j (\bra{\gamma}{\hat n_j}\ket{\gamma}-\bra{\alpha}{\hat n_j}\ket{\alpha})e_j}$
due to mean field is of order $\mathcal{O}(\lambda^2)$.
As mentioned above, since the correlation function will enter the linear response formula for fidelity \cite{gorinNJP6}
with a prefactor $\lambda^2$,  the overall correction to the fidelity amplitude is $\mathcal{O}(\lambda^4)$
and will be neglected.
We split the $C_\alpha(\tau)$
into three groups according to the classification of two-body transitions
\begin{eqnarray}
C_\alpha(\tau)
&=& \langle \alpha| v^{-2} \sum_{kl} c_k^\dagger c_l^\dagger c_l c_k c_k^\dagger c_l^\dagger c_l c_k  \nonumber \\
&+& v^{-2}\sum_{i\neq k,j}  c_k^\dagger  c_j^\dagger c_j c_i c_i^\dagger c_j^\dagger c_j c_k \ee^{-2\pi\ii (e_i-e_k)\tau}  
\label{eq:spectralC} \\
&+& v^{-2}\!\!\!\sum_{i<j \neq k<l}%
\!\!\!c_k^\dagger c_l^\dagger c_j c_i c_i^\dagger
c_j^\dagger c_l c_k \ee^{-2\pi\ii (e_i+e_j-e_k-e_l)\tau} |\alpha\rangle . \nonumber
\end{eqnarray}
The abbreviation in the last sum means that indices $i,j,k,l$ are all different.
The first term of~(\ref{eq:spectralC}) is actually only present if we would be interested
in considering the full two-body term as perturbation and thus consider the corresponding dephasing
 \footnote{If we wish to consider a time reversal
invariant system, the two-body coefficients $V_{ijkl}$
can be chosen real with variances
$\langle V_{ijkl}V_{mnop}\rangle_V = v^{-2}(\delta_{ij,op}\delta_{kl,mn}+\delta_{ij,mn}\delta_{op,kl})$, and as
a result, the first, time-independent term in~(\ref{eq:finalCalpha})
should be multiplied by factor 2;
obviously corresponding single particle levels should be chosen.}.
Partially summing the expression yields
\begin{eqnarray}
       C_\alpha(\tau) &=& v^{-2}{M\choose 2}
       + v^{-2}(M-1)\sum_{ik} n_k{\bar n_i}\ee^{-2\pi\ii(e_i-e_k)\tau}\nonumber\\
       &+& v^{-2}\frac{1}{4}\sum_{ijkl}{\bar n_i}{\bar n_j}n_k n_l \ee^{-2\pi\ii(e_i+e_j-e_k-e_l)\tau}
       \label{eq:finalCalpha}
\end{eqnarray}
Here $n_i\!=\!\langle \alpha | c_i^\dagger c_i | \alpha\rangle$ indicates the
Fermion occupation number
and we use the convention ${\bar n_i}\!=\!1\!-\!n_i$.
The relation between the fidelity amplitude and the correlation function after averaging over the
interaction reads
\begin{equation}
       \langle f(t) \rangle_V = \sum_\alpha |x_\alpha|^2 \Big[
               1-4\pi^2 \lambda^2 \int_0^t\!\dd t_1 \int_0^{t_1}\!\dd t_2 C_\alpha(t_1-t_2)\Big].
       \label{eq:fidspecconnection}
\end{equation}

If the spectrum of $H_0$ is known, this allows us to
determine the fidelity amplitude by integrating the
correlation function in~(\ref{eq:spectralC}) twice for each eigenstate $\alpha$ and
evaluate the sum in~(\ref{eq:fidspecconnection}). Actually we are more interested in ensemble behavior
which we shall study in the next two subsections.

\section{Ensemble averaged fidelity amplitude}
\label{sec:ensembleavg}
Relation~(\ref{eq:fidspecconnection}) can be simplified
if we average over initial states or select a random initial state by choosing
Gaussian random coefficients $x_\alpha$ normalized by the requirement
$\langle |x_\alpha|^2 \rangle_0=1/\mathcal{N}$.
The correlation function can be averaged as
$C(\tau) = \frac{1}{\mathcal{N}} \sum_\alpha \langle C_\alpha(\tau)\rangle_0$
which implies averaging over all distributions of $M$ particles on $N$ orbitals.
%
We find
\begin{eqnarray}
       C(\tau) &=& \frac{v^{-2}}{\mathcal{N}} (M-1) {N-2\choose M-1}
               \sum_{m\neq n} \langle \ee^{-2\pi\ii(e_m-e_n)\tau}\rangle_0 + \nonumber \\
         &+&     \frac{ v^{-2} }{\mathcal{N}} {N-4\choose M-2}\!\!%
        \sum_{m\neq n \neq p \neq q}\!\!\!\!\!\!\!\langle\ee^{-2\pi\ii (e_m+e_n-e_p-e_q)\tau}\rangle_0. \label{eq:spectralC0}
\end{eqnarray}
A closer inspection of first term in~(\ref{eq:spectralC0}) shows that it is connected
with the probability distribution of pairs of energies of two \emph{arbitrary}
orbitals which is given by Dyson's \emph{two-point correlation function} $R_2$
following the definition in \cite{mehtabook}. The average can then be obtained by integration.
A similar argument connects the
second average with the four-point correlation function $R_4$. Finally, using the
normalization factor $v^2={M\choose 2}\big(2(N-M)+{N-M \choose 2}\big)$
we rewrite the correlation function in compact form 
(using $\eta=\frac{N-M-1}{4}$)
\begin{eqnarray}
       C(t) &=& \frac{1}{1+\eta}\mathcal{F}[R_2]\left(\frac{2\pi\sqrt{2} t}{d}\right)  + \nonumber \\
       &+&
	   \frac{1}{1+\eta^{-1}} \mathcal{F}[R_4]\left(\frac{2\pi\sqrt{2} t}{d}\right).
       \label{eq:spectralc0final}
\end{eqnarray}
The functions $\mathcal{F}[R_n](t)$ are obtained by Fourier
transformation of the correlation functions $R_n$ (see Appendix 
for precise definitions). In the case of GUE
single-particle spectra, the functions $\mathcal{F}[R_n](t)$ can be calculated
analytically and are given as products of finite polynomials in $t$ and
Gaussians (see Appendix).
They are normalized such that $\mathcal{F}[R_{2,4}](0)=1$.

The simulations have been performed both including and excluding the diagonal terms 
of the two-body interaction and coincide with the theoretical prediction
in~(\ref{eq:spectralc0final}) (or the corresponding full calculation)
if the random states are chosen from the whole energy spectrum or from its center.
The ensemble and state averaged fidelity amplitude can now be obtained by integrating
the averaged correlation function in~(\ref{eq:spectralc0final}) twice
\begin{equation}
       \langle f(t)\rangle= 1-4\pi^2 \lambda^2 
       \int_0^t \dd t_1 \int_0^{t_1}\dd t_2 C(t_1-t_2).
       \label{eq:spectralfidelity}
\end{equation}
In the {\em residual case} where the perturbation $V$ has no diagonal terms (i.e. in the case of
mean field unperturbed Hamiltonian) the correlation function has no 
constant term. The correlation integral can then be approximated by a linear 
function for fairly long times as long as the linear response formalism is 
justified. The simulation (Fig.~\ref{fig:fidth}) is done with random initial 
states from the center of the spectra and is in agreement with the
full-spectra prediction. The residual case thus coincides qualitatively
with the result of a general residual perturbation
of a Hamiltonian with a random spectrum \cite{G-prl,fidelityreview}
also known as POE \cite{dittes}. Fidelity freeze is not found and we could easily 
conclude that this exercise is rather disappointing. Yet, considering the known non-ergodicity of the TBRE,
it seems worthwhile to check whether the average behavior reflects the typical behavior.
\begin{figure}
       \centering
       \includegraphics[width=8cm]{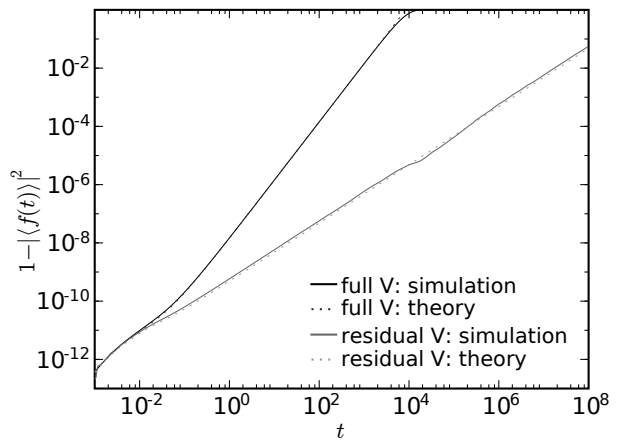}
       \caption{Fidelity amplitude decay for $6$ particles in $12$ orbitals averaged over
       $2\cdot 10^5$ realizations of $V$, $H_0$ and initial states,
       with the perturbation strength $\lambda=10^{-4}$.
       For sufficiently long times the decay is quadratic in the case
       of full perturbation (upper) and linear in the residual case (lower).
       Initial states are taken from the center of spectra. 
       }
       \label{fig:fidth}
\end{figure}

\section{Fidelity decay for a typical ensemble member}
\begin{figure}
       \centering
       \includegraphics[width=8cm]{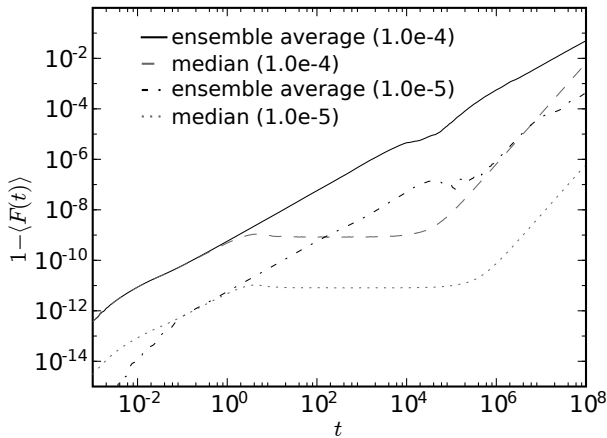}
       \caption{ Average fidelity in $10^{5}$ realizations of $H_0$
       and $V$ compared to the median fidelity for $6$
       particles in $12$ orbitals and with two values of perturbation strengths $\lambda=10^{-4}$
       and $\lambda=10^{-5}$.
       The median fidelity decay (long-dashed, dotted) shows a plateau practically absent 
       in the ensemble average (full, short-dashed curves).}
       \label{fig:median}
\end{figure}
        We therefore proceed to analyze the median of fidelity decay.
For this purpose we use fidelity $F(t)=|f(t)|^2$, which is a real quantity, instead of fidelity 
amplitude $f(t)$. (We note however that the results are practically the same for the
real part of fidelity amplitude.)
We define
\emph{median fidelity} $F_\textrm{m}(t)$ such that at any time
$F_\textrm{m}(t)$ is lower than fidelity $F(t)$
in half of the realizations.
For a randomly chosen member of the ensemble
there is thus a $0.5$ probability that a plateau in fidelity decay
will last longer than the median fidelity plateau (Fig.~\ref{fig:median}).

A useful alternative in such situation is to consider the average
of the logarithm of the quantity under consideration -- in our case
$\langle \ln(1-F(t)) \rangle$. This quantity has been used in elastodynamics under
the name distortion as we have discussed in the introduction.
The distortion shown in figure~\ref{fig:logmedian} in the
following section has essentially the same feature as the median.
Note that it is not obvious how to obtain analytical results
for either of these quantities, but the latter may be slightly more accessible.

\begin{figure}
       \centering
       \includegraphics[width=8cm]{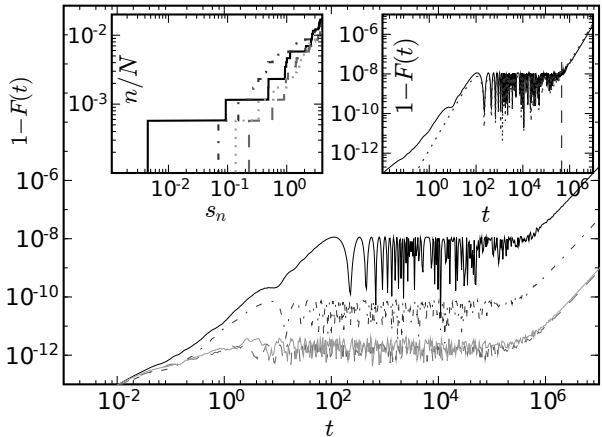}
	   \caption{ Fidelity 
       decay in a few individual ensemble members for
       the residual case with $6$ particles and $12$ orbitals. The upper left inset
       shows a small region of the corresponding cumulative level spacing
       distribution of the unperturbed spectra $H_0$ (We plot a relative number of level spacing 
       $n/{\cal N}$ below $s_n$.) The agreement with the
       minimum spacing only theory (dotted) is shown in the upper right inset, 
	   dashed line denotes the ending time $t_\textrm{e}$ (eq. (\ref{eq:te}).}
       \label{fig:explain}
\end{figure}
It remains to be understood how the freeze, which
is present in most realizations, comes about and is eventually completely averaged out.
To elucidate this problem fidelity of a few individual
ensemble members is presented (Fig.~\ref{fig:explain}). However, the
cases shown in the figure are not all equally probable; the upper one
occurs rarely whereas the lower two curves truly represent the majority of cases.
The upper left inset of Fig.\ref{fig:explain} shows the beginning of
the cumulative level spacing distribution of $H_0$ corresponding to
the fidelity curves shown in the main figure.
Interestingly, the plateau properties of the exceptional large deviation cases with high plateaus
are determined by the minimum energy spacing alone and the
conjecture is that the plateau vanishes together with the smallest level spacing.

Another peculiarity
results from the two-body nature of interaction which only connects certain levels as
shown in section~\ref{sec:ensembleavg}.
Transitions involving three orbitals will very likely have big energy difference
because of the level repulsion in single-particle spectra; this contrasts with
the transitions involving four orbitals that can easily have small energy differences
if one particle lowers the energy by roughly the same amount by which the other one raises it.
The effect of a such two-body operator will be greater than the effect of any other operator.

Considering the established connection between the level spacing of four-orbital two-body
operators and the fidelity amplitude plateau, an illustration is made where
all three-orbital terms are omitted. Each four-orbital two-body operator
$c_i^\dagger c_j^\dagger c_l c_k$ is
in a unique way connected with the energy difference $\mu_{ijkl}:=e_i\!+\!e_j\!-\!e_k\!-\!e_l$ 
in the spectrum of $H_0$
and the four-orbital part of
the correlation function~(\ref{eq:spectralC})
averaged over the perturbation becomes a sum over all possible positive spacings
\begin{eqnarray}
       C_\alpha(t) &=&  v^{-2}\!\!\!\sum_{i<j \neq k<l}\Big[%
	   \bra{\alpha}{\bar n_i}{\bar n_j}n_ln_k\ee^{-2\pi\ii|\mu_{ijkl}|}\ket{\alpha}\nonumber \\
	 &+& \bra{\alpha}{\bar n_k}{\bar n_l}n_jn_i\ee^{2\pi\ii|\mu_{ijkl}|}\ket{\alpha}\Big].
\end{eqnarray}
Assuming random initial states averaging can be performed
\begin{equation}
       \frac{1}{\mathcal{N}}\sum_\alpha C_\alpha(t) =
       \frac{ 8 (N-4)!}{N! (1+\frac{4}{N-M-1})} \sum_{q} \cos 2\pi\mu_q t
\end{equation}
and the fidelity amplitude is obtained by integration~(\ref{eq:spectralfidelity}) as
\begin{equation}
       f(t) = 1 - \lambda^2 \frac{ 8 (N-4)!}{N! (1+\frac{4}{N-M-1})}
       \sum_{q} \frac{1-\cos 2\pi\mu_q t}{\mu_q^2}.
\end{equation}
If the smallest degenerate spacing $\mu_0$ is orders of magnitude smaller than any other
spacing, the sum above can be approximated by the smallest spacing term alone.
All other terms are smaller by a factor $(\mu_0/\mu_q)^2$. This approximation
is very illustrative and
can be used to estimate the beginning time $t_{\rm b}$ and the shift
of the plateau by the condition $\mu_0 t_{\rm b} = 1$ and by removing
the time-dependent part, respectively. Estimation of the ending
time $t_{\rm e}$ of the plateau is a little more tedious. There we need the fourth-order terms in
the linear response formula~(\ref{eq:linresponse}). Following the same principles
and keeping the minimum spacing term only
we obtain the fourth-order correction to fidelity
of highest order in time
(we are interested in very long times) as
\begin{equation}
	f^{(4)}(t)=\frac{128\pi^2 (N-4)!(N\!-\!M\!-\!1)}{N!(N\!-\!M)M(M\!-\!1)(N\!-\!M\!+\!3)^2} \frac{\lambda^4 t^2}{\mu_0^2}.
\end{equation}
The ending time $t_{\rm e}$ can then be estimated by comparing amplitudes of the
second and fourth-order terms which gives the well known\cite{freezeint,freezech}  $\lambda^{-1}$ dependence
\begin{equation}
       t_{\rm e} = \frac{1}{2\pi\lambda} \sqrt{{M\choose 2}(N-M)(N-M+3)}.
       \label{eq:te}
\end{equation}
The ending time does not depend on level spacings, which is
again in agreement with the data shown in Fig.~\ref{fig:explain}.
If $t_{\rm b} \sim t_{\rm e}$ the plateau, which is a pure second-order
phenomenon, would begin in the region where the second order approximation is
no longer valid and hence cannot exist.
The upper right inset on Fig.~\ref{fig:explain} shows that the second and fourth
order terms together describe all important features of the fidelity amplitude
decay for the uppermost realization in Fig~\ref{fig:explain}
with an extremely small level spacing.

We now understand why the plateau, nearly always present
in individual realizations, does not appear in the ensemble average.
Realizations of the unperturbed spectrum with an extremely small level spacing occur rarely but
when averaging the fidelity amplitude they eventually dominate the behavior and
the plateaux are averaged out.

\section{Comparison to other models}
\begin{figure*}
       \centering
       \includegraphics[width=8cm]{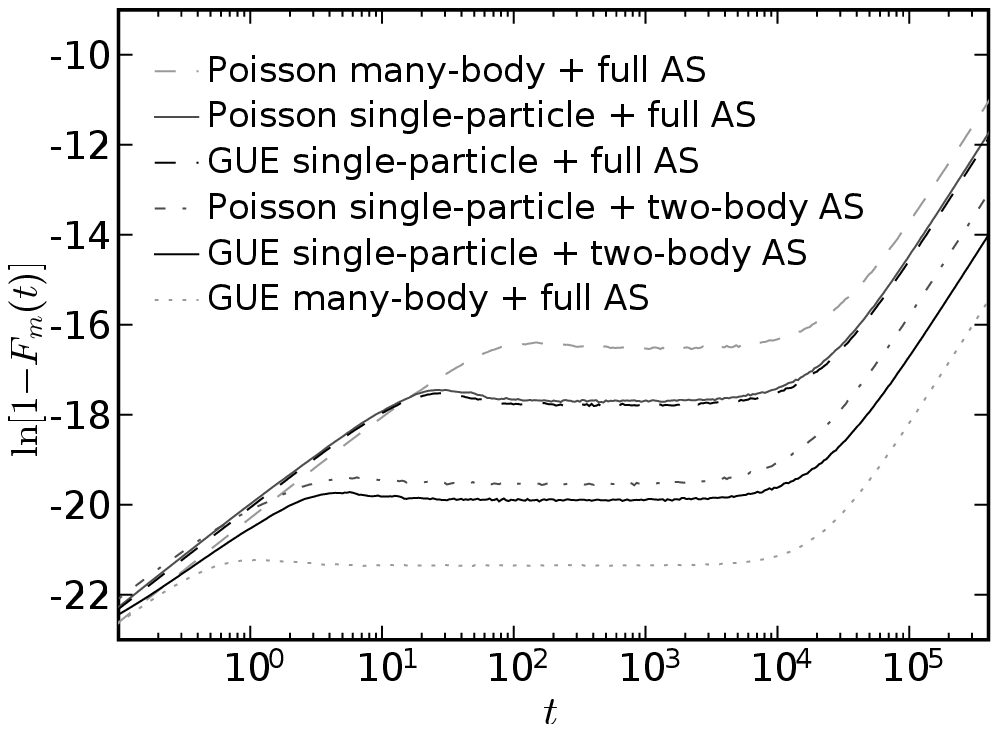}%
       \includegraphics[width=8cm]{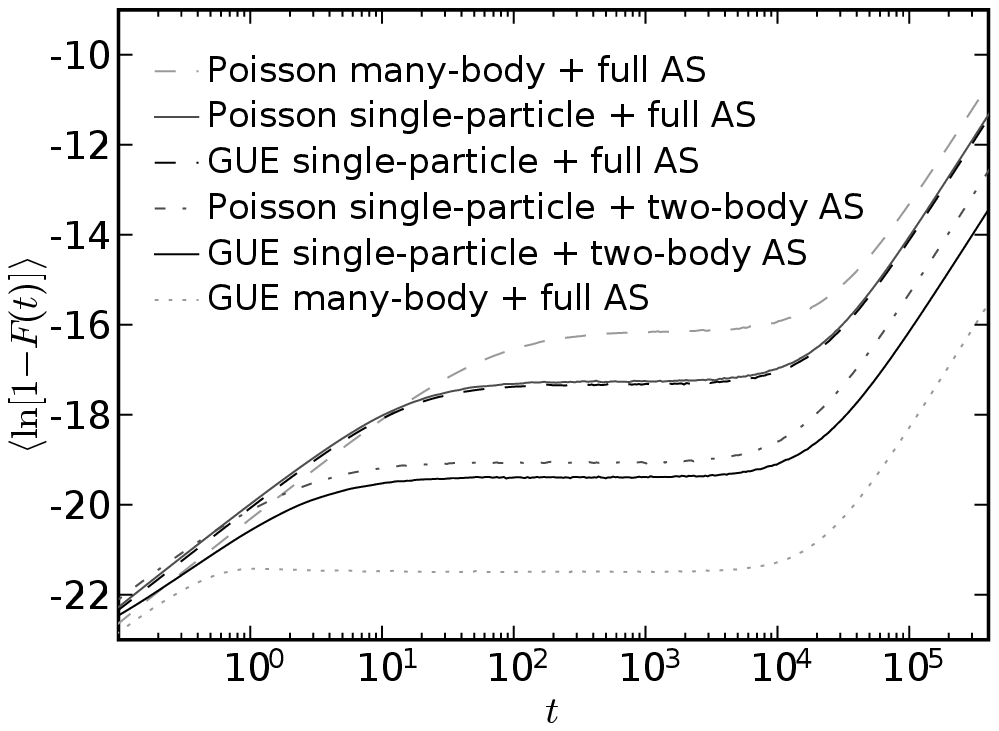}
       \caption{Median fidelity decay (left)  and average natural logarithm of $1-F(t)$ (right) for
       various models (described in text and labeled in the figure - ordered from top to bottom)
       for $5$ particles in $10$ orbitals and
       perturbation strength $\lambda=10^{-4}$. Perturbation is always imaginary antisymmetric.
       The results are averaged over $2 \cdot 10^{4}$ realizations.}
       \label{fig:logmedian}
\end{figure*}
The theory we have developed to describe the behavior of the
model under consideration can be applied to other physical models
and used to understand the differences between them.
The key features of the model (\ref{eq:H}) are 1) level repulsion in single particle spectra and
2) sparse perturbation.
By comparing to models with
full random matrix spectra which correspond to chaotic dynamics on the one hand,
and independent Poissonian spectra which corresponds to regular dynamics
on the other hand, we find that our model lies somewhere in between as we can see from
the Fig.~\ref{fig:logmedian}.
Besides the median fidelity which corresponds to the typical behavior,
an alternative measure namely the average of the logarithm of $1-F(t)$ (also known a distortion) is considered
in parallel. Although not completely identical, they both show essentially the same features and
to some extent even quantitative agreement.
In order to eliminate any potential effect of diagonal perturbation terms,
we consider in the following models always purely imaginary antisymmetric perturbation,
individually normalized in the usual way
(${\rm tr}[V^2]=\mathcal{N}$). However, we will be switching between the cases of a \emph{full} and
a \emph{two-body} perturbation matrix.

The first model under consideration (uppermost) has a random spectrum of $H_0$ such that
the level spacing distribution is Poissonian. The perturbation is chosen
from an ensemble of \emph{full} antisymmetric matrices. The Poissonian level spacing distribution favors
small spacings between spectral levels which contribute to fast (initially linear) fidelity decay.
In this case the type of perturbation has no significant effect as small spacings
are probable for any pair of levels. This is not true for $H_0$ constructed from 
independent single particle spectra which we take as
 Poissonian or GUE as in our original model (\ref{eq:H}).
For the latter, small level spacing for \emph{some} transitions (e.g. three-orbital
two-body transitions) is very improbable. If the perturbation is a full matrix, the effect of
such transitions is relatively small and there is no significant difference between
the two cases. If, on the other hand, only a few levels are coupled as in the case of
two-body interactions, three-orbital transitions in the
case of Poissonian single particle spacings \emph{can} and in the case of GUE spacings \emph{cannot} involve
small spacings. Thus in the former case they cannot be neglected. Typical decay of fidelity is in both cases
slower than in the case of full perturbation.
For illustration, also a completely different physical situation with
random matrix many-body spectrum of $H_0$ is shown. Because of the
level repulsion in many-body spectra, \emph{any} transition will very unlikely
have small level spacing. Only then, the plateau in fidelity decay is present not only in a typical case but
also on average.

\section{Ground state fidelity amplitude}
\begin{figure}
       \centering
       \includegraphics[width=8cm]{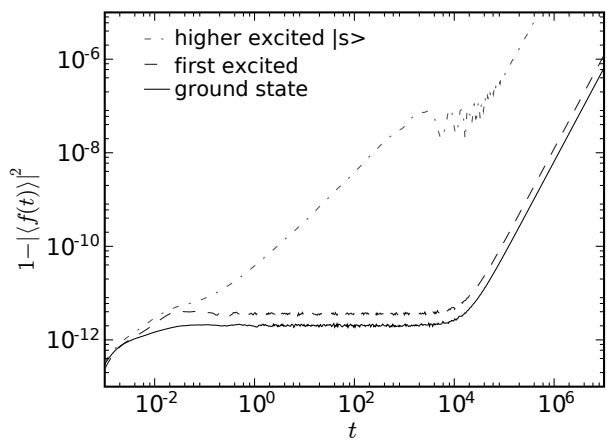}
       \caption{ Fidelity amplitude decay for the ground state and first excited state
       compared to higher excited state $\ket{s}$ (eq. (\ref{eq:higher})) for
	   $6$ particles in $12$ orbitals and perturbation strength $\lambda=10^{-4}$.
	   The results are averaged over $10^{4}$ realizations of
       perturbation and spectra.}
       \label{fig:ground}
\end{figure}
We have up to now considered only random initial states, but the behavior of the
 ground state of the  independent particle model $H_0$, i.e. the Hartree-Fock  ground state,
under a perturbation formed by residual interactions is also of interest.
The correlation function can now no longer be simplified by state averaging.
The main problem is to determine
the matrix element $[V(t)V(0)]_{gg}=C_g(t)$ where
$\ket{g}=c_{0}^\dagger\!\cdots\!c_{M-1}^\dagger\ket{0}$ is the ground state
occupying the lowest $M$ single-particle levels. Regardless of the realization
of the spectra of $H_0$, any non-diagonal two-body operator raises the energy
for at least the minimum level spacing in the single-particle spectra where
levels are unlikely to be close together.
The same incidentally holds in the first excited state with the only difference that there
the energy can also be lowered by the same amount.
Note that more freedom does exist for higher excited states such as for example
\begin{equation}
\ket{s}=c_0^\dagger\!\cdots\!c_{M-2}^{\dagger}c_M^{\dagger}c_{M+1}^\dagger\ket{0}
\label{eq:higher}
\end{equation}
where the operator $c_{M-1}^{\dagger}c_{M+2}^{\dagger}c_{M+1}c_{M}$ or
its hermitian conjugate
increases or decreases energy by $e_{M-1}+e_{M+2}-e_M-e_{M+1}=(e_{M+2}-e_{M+1})-(e_M-e_{M-1})$
which indeed \emph{can} be very small.

This leads to the suspicion that the freeze might exist on average for these two lowest independent
particle initial states.
The numerical results indeed show (Fig.~\ref{fig:ground}) the plateau in the ensemble averaged
fidelity amplitude for the ground and the first excited state.

\section{Conclusion}
We have analyzed a very elementary many-fermion model in context of
echo dynamics. The model describes spinless fermions whose underlying
single-particle dynamics is chaotic and are perturbed by
random two-body interactions. Our interest was focused on the stability
of a mean field approximation where the diagonal terms of the interaction
have been included in the unperturbed Hamiltonian.
For weak perturbations the  decay of the fidelity amplitude in this case
typically displays
a high-level plateau also known as freeze of fidelity.
This freeze lasts for times long on the scale of the
Heisenberg time.
The unexpected point is that the freeze typically present in most realizations
of members of the ensemble will vanish on average giving a very dramatic
example of the non-ergodicity of the TBRE. To see the effect which should
dominate most experiments and should hence be observable, we have to
consider median behavior or consider
logarithmic averages. This fact beyond the specific interest for the TBRE
could pave the road for interesting new applications of random matrix theory
analyzing non-ergodic situations.

\begin{acknowledgments}
IP and TP acknowledge hospitality of CiC (Cuernavaca), where this work
was initiated, and financial support by the grants P1-0044 and J1-7347 of Slovenian 
Research Agency. THS acknowledges support from the CONACyT grant 44020-F
and UNAM - PAPIIT grant IN112507.\\

\end{acknowledgments}

\appendix* \section{Analytical calculation of the correlation function}
\label{app:correlation}
The final compact expression for the averaged correlation
function $C(\tau)$ (see eq.~\ref{eq:spectralc0final}) involves
Fourier integrals of the Dyson's correlation functions  $R_n(\mathbf{x})$
\begin{equation}
       \mathcal{F}[R_n](k)=\frac{(N\!-\!n)!}{N!}\int_{-\infty}^\infty\!\!\!R_n(\mathbf{x})
       \ee^{-\ii k\sum_{j=0}^{n-1}(-1)^j x_j}\dd^n\mathbf{x}.
       \label{eq:correlationF}
\end{equation}
If the correlations functions $R_n$ correspond to GUE,
such integrals can be analytically integrated and presented in terms of
functions $J_{nm}(t)$:
\begin{equation}
       J_{nm}(t) = \int_{-\infty}^\infty H_n(x) \ee^{-x^2} H_m(x)\ee^{-\ii t x} \dd x
\end{equation}
where $H_n(x)$ are standard Hermite polynomials. It can easily be shown using
recursion that a function $J_{mn}(t)$ can be written as finite series
\begin{equation}
	J_{mn}(t)=\frac{\ee^{-t^2/4} (-\ii t)^{n+m}}{\sqrt{ 2^{n+m}n!m!}}%
	\sum_{j=0}^{\textrm{min}(m,n)}\!\!\!(-2)^j j!{m\choose j}{n\choose j}\frac{1}{t^{2j}}.
\end{equation}
The simplest case is the transformation of
$R_1(x)=\sum_{n=0}^{N-1} \frac{1}{2^n n! \sqrt{\pi}} \ee^{-x^2}H_n(x)^2$
which gives a simple result
\begin{equation}
       \mathcal{F}[R_1](t) = \sum_{n=0}^{N-1} J_{nn}(t).
\end{equation}
Higher level correlation functions $R_n$ are expressible by lower level
$R_n$ and cluster functions $T_2$ \cite{mehtabook}.
Because $\mathcal{F}$ is a linear transformation it can also be expressed
by lower order transformations where transformations of cluster functions
involve terms without the sign factors $(-1)^j$ in the exponential in~(\ref{eq:correlationF}),
denoted by $\mathcal{F}[T_n]^+$.
Straight-forward calculation gives
\begin{eqnarray}
	\mathcal{F}[T_2]^{\mp}(t) &=& \sum_{nm} J_{nm}(t)J_{nm}(\mp t) \\
       \mathcal{F}[T_3]^{\mp}(t) &=& \sum_{nmk} J_{mn}(t) J_{mk}(-t)J_{nk}(-t) \\
       \mathcal{F}[T_4]^{\mp}(t) &=& \sum_{mnkl} J_{ml}(t) J_{mn}(t)J_{kl}(-t)J_{nk}(-t).\quad\quad
\end{eqnarray}
Finally, following the equality $R_2(x,y)=R_1(x)R_1(y)-T_2(x,y)$ and
similarly for $R_4$ it finally holds
\begin{eqnarray}
\mathcal{F}[R_2](t) &=& \mathcal{F}[R_1](t)^2 - \mathcal{F}[T_2]^-(t)\label{eq:FspecR2} \\
\mathcal{F}[R_4](t) &=& \mathcal{F}[R_1](t)^4 + \mathcal{F}[T_2]^+(t)^2 +
       2 \mathcal{F}[T_2]^-(t)^2 \nonumber \\
	   &-& 2 \mathcal{F}[R_1](t)^2 \mathcal{F}[T_2]^+(t) - 4 \mathcal{F}[R_1](t)^2 \mathcal{F}[T_2]^-(t) \nonumber \\
	   &+& 8 \mathcal{F}[R_1](t)\mathcal{F}[T_3](t) - 6 \mathcal{F}[T_4](t).
       \label{eq:FspecR4}
\end{eqnarray}
Hence, the averaged correlation function, although not as a compact expression,
can be analytically evaluated using the above equations.


\begin{thebibliography}{10}

\bibitem{wigner1}
E.~P. Wigner,
\newblock Ann. Math. {\bf 53}, 36 (1951);
E.~P. Wigner,
\newblock Proc. Cambridge Philos. Soc. {\bf 47}, 790 (1951).

\bibitem{cartan}
E.~Cartan,
\newblock Abh.\ Math.\ Sem.\ Univ.\ Hamburg {\bf 11}, 116 (1935).

\bibitem{brody}
T.~Brody {\em et~al.},
\newblock Rev. Mod. Phys. {\bf 53}, 385 (1981).

\bibitem{balian}
R.~Balian,
\newblock Nuovo Cimento {\bf B57}, 183 (1958).

\bibitem{guhr}
T.~Guhr, A.~M{\"u}ller-Groeling, and H.~A. Weidenm{\"u}ller,
\newblock Phys.\ Rep. {\bf 299}, 189 (1998).

\bibitem{frenchPL33B}
J.~French and S.~Wong,
\newblock Phys. Lett. {\bf 33B}, 449 (1970).

\bibitem{bohigasPL34B}
O.~Bohigas and J.~Flores,
\newblock Phys. Lett. {\bf 34B}, 261 (1971).

\bibitem{french2}
J.~B. French and S.~S.~M. Wong,
\newblock Phys. Lett. {\bf 35B}, 5 (1971).

\bibitem{benet}
L.~Benet and H.~A. Weidenm{\" u}ller,
\newblock J. Phys.~A: Math. Gen. {\bf 36}, 3569 (2003).

\bibitem{bertsch1}
C.~W. Johnson, G.~F. Bertsch, and D.~J. Dean,
\newblock Phys. Rev. Lett. {\bf 80}, 2749 (1998);
C.~W. Johnson, G.~F. Bertsch, D.~J. Dean, and I.~Talmi,
\newblock Phys. Rev. C {\bf 61}, 014311 (1999).

\bibitem{papenbrockweiden}
T.~Papenbrock and H.~A. Weidenm{\" u}ller,
\newblock Nucl. Phys. A {\bf 575}, 422 (2005).

\bibitem{frank}
R.~Bijker, A.~Frank, and S.~Pittel,
\newblock Phys. Rev. C {\bf 60}, 021302 (1999).


\bibitem{french3}
J.~B. French,
\newblock Rev. Mex. Fis. {\bf 22}, 221 (1973).

\bibitem{zelevinsky}
V.~Zelevinsky, B.~A. Brown, N.~Frazier, and M.~Horoi,
\newblock Phys. Rep. {\bf 276}, 85 (1996).

\bibitem{floresseligman}
J.~Flores, M.~Horoi, M.~{M\" uller}, and T.~H. Seligman,
\newblock Phys. Rev. E {\bf 63}, 026204 (2000).

\bibitem{kota}
V.~K.~B. Kota,
\newblock Phys. Rep. {\bf 347}, 223 (2001).

\bibitem{peresPRA30}
A.~Peres,
\newblock Phys. Rev. A {\bf 30}, 1610 (1984).

\bibitem{nielsenbook}
M.~A. Nielsen and I.~L. Chuang,
\newblock {\em Quantum Computation and Quantum Information} (Cambridge
  University Press, Cambridge, 2000).
  
 \bibitem{fidelityreview}
T.~Gorin, T.~Prosen, T.~H. Seligman, and M.~{\v Z}nidari{\v c},
\newblock Phys. Rep. {\bf 435}, 33 (2006).

\bibitem{jalabert}
R.~F.~Jalabert and H.~M.~Pastawski, Phys. Rev. Lett. {\bf 86}, 2490 (2001).

\bibitem{lewenkopf}
F.~Cucchietti, C.~Lewenkopf, E.~Mucciolo, H.~Pastawski and R.~Vallejos, Phys. Rev. E {\bf 65},
046209 (2002).

\bibitem{cucchietti}
F.~M.~Cucchietti, D.~A.~R. Dalvit, J.~P.~Paz, and W.~H.~Zurek, 
Phys. Rev. Lett. {\bf 91}, 210403 (2003).

\bibitem{vanicek}
J.~Vanicek and E.~Heller, Phys. Rev.E {\bf 68}, 056208 (2003);
J.~Vanicek, Phys. Rev. E {\bf 70}, 055201(R), (2004); {\it ibid.} {\bf 73}, 046204 (2006).

\bibitem{stoeckmannNJP}
H.-J.~St\" ockmann and R.~Sch\" afer,
New J. Physics {\bf 6}, 199 (2004).

\bibitem{stoeckmannPRE73}
H.-J. St{\"o}ckmann and H.~Kohler,
\newblock Phys. Rev. E {\bf 73}, 066212 (2006).

\bibitem{gorinNJP6}
T.~Gorin, T.~Prosen, and T.~H. Seligman,
\newblock New J. of Physics {\bf 6}, 20 (2004).

\bibitem{P02}
T.~Prosen,
\newblock Phys. Rev. E {\bf 65}, 036208 (2002).

\bibitem{PZ02}
T.~Prosen and M.~{\v Znidari\v c},
\newblock J. Phys. A {\bf 35}, 1455 (2002).

\bibitem{Progress}
T.~Prosen, T.~H. Seligman, and M.~{\v Z}nidari{\v c},
\newblock Prog. Theor. Phys. Suppl. {\bf 150}, 200 (2003).

\bibitem{G-prl}
T.~Gorin {\it et al.},
\newblock Phys. Rev. Lett. {\bf 96}, 244105 (2006).

\bibitem{jacquodPRE}
Ph.~Jacquod, P.~G.~Silvestrov, and C.~W.~J.~Beenakker, Phys. Rev. E {\bf 64},
055203(R) (2001).

\bibitem{cerruti}
N.~R.~Cerruti and S.~Tomsovic, Phys. Rev. Lett. {\bf 88}, 054103 (2002).

\bibitem{freezeint}
T.~Prosen and M.~{\v Z}nidari{\v c},
\newblock New J. Phys. {\bf 5}, 109 (2003).

\bibitem{freezech}
T.~Prosen and M.~{\v Z}nidari{\v c},
\newblock Phys. Rev. Lett. {\bf 94}, 044101 (2005).

\bibitem{weaver}
O.~I. Lobkis and R.~L. Weaver,
\newblock Phys. Rev. Lett {\bf 90}, 254302 (2003).

\bibitem{gorin-weaver}
T.~Gorin, T.~H. Seligman, and R.~L. Weaver,
\newblock Phys. Rev. E {\bf 73}, 015202(R) (2006).

\bibitem{leyvraz-privat}
F.~Leyvraz, 2006,
\newblock private communication.

\bibitem{flambaum}
V.~V.~Flambaum, G.~F.~Gribakin, and F.~M.~Izrailev,
Phys. Rev.~E {\bf 53}, 5729 (1996);
V.~V.~Flambaum and F.~M.~Izrailev,
Phys. Rev.~E {\bf 61}, 2539 (2000).

\bibitem{alhassid}
Y.~Alhassid, Ph.~Jacquod, and A.~Wobst,
\newblock Phys. Rev.~B {\bf 61}, R13357 (2000).

\bibitem{dittes}
F.-M. Dittes, I. Rotter, and T. H. Seligman, 
Phys. Lett. A {\bf 158}, 14 (1991).

\bibitem{mehtabook}
M.~L. Mehta,
\newblock {\em Random Matrices (Revised and Enlarged, 2nd Edition)} (Academic
  Press, London, 1991).

\end{thebibliography}
\end{document}